# Theoretical Considerations on Compensation of the Accommodation-Vergence Mismatch by Refractive Power of Focus-Adjustable 3D Glasses

Dal-Young Kim*

Department of Optometry, Seoul National University of Science and Technology (Seoul Tech), Seoul 139-743, Korea

* Corresponding author: dykim@seoultech.ac.kr

ABSTRACT

The accommodation-vergence mismatch has been considered as a cause of the visual fatigue induced in watching 3D display. We would propose the mismatch can be compensated by refractive power of adjustable-focus 3D glasses. From lens optics and geometrical considerations, we also developed equations that calculate required refractive power of the 3D glasses. The compensation we proposed is supposed to reduce the visual fatigue of 3D display.

**Introduction**

Owing to rapid growth of three-dimensional (3D) TV and cinema industry, the visual fatigue problem of 3D display is drawing much attention. Although some researchers recently claimed that the accommodation-vergence mismatch (AVM) was not an essential cause [1,2], the AVM has been regarded as a main cause of the 3D visual fatigue [3-5]. A lot of ideas have been proposed to solve it [6], most of which are still under development.

Here we would propose another solution based on optometry, giving some theoretical considerations on it.

**Theoretical Background**

A simple diagram of Fig. 1(a) depicts that the viewer is watching a two-dimensional (2D) display. $N_1$ and $N_2$ are the nodal points of left and right eyeballs. Visual axes of the viewer's both eyes converge to a fixation point (FP) on the 2D display with vergence angle of $\theta_1$. Object lengths are $N_1$-FP and $N_2$-FP for left and right eye, respectively. (For simplicity, hereafter we will consider the left eye only. It is because the same considerations can be applied to the right eye.) The eye accommodation is defined as an inverse of the object length, so that accommodation of the left eye becomes $1/(N_1\text{-FP})$.

On the other hand in Fig. 1(b), the viewer is watching a 2-view type 3D display, in which every frame of displayed motion pictures is divided into two pictures ($FP_1$ and $FP_2$) for stereoscopic vision. One picture is incident from $FP_1$ to the left eye while the other is incident from $FP_2$ to the right eye. As a result, artificial binocular disparity is induced, locating a virtual 3D image at V point with changed vergence angle of $\theta_2$. Although distance from the left eye to the virtual image is $N_1$-V ($\equiv d_V$), real object distance is $N_1$-$FP_1$ ($\equiv d_O$). It is because displayed real picture is not at V but at $FP_1$. Likewise, accommodation of the left eye is not $1/d_V$ but $1/d_O$.

It is well known that the accommodation and vergence angle are not independent but coupled with each other [7]. Strong accommodation and wide vergence angle must come together when object is at near point, while weak accommodation and narrow vergence angle do when object is at far point. The viewer feels visual discomfort if this coupling is broken. Long object distance $d_O$ (that is, weak accommodation $1/d_O$) with wide vergence angle $\theta_2$ as depicted in Fig. 1(b) is quite unusual, and cause visual discomfort. This is the AVM that means a mismatch between viewer's accommodation and binocular vergence angle when he/she watches 3D display or head-mounted display.

**Compensation by Refractive Power of 3D Glasses**

Fig. 2(a) describes the AVM simplified in a viewpoint of the object distances. If we can change the eye accommodation from $1/d_O$ ($\equiv D_O$) to $1/d_V$ ($\equiv D_V$) keeping focus on $FP_1$, the accommodation-vergence coupling (A-V coupling) would be kept unbroken, so that the visual discomfort could be removed. It can be achieved by adding extra refractive power of an eye-glasses lens.

For watching the 2-view type 3D display in real, the viewer has to wear 3D glasses as shown in Fig. 2(b). Up-to-date conventional 3D glasses do not have refractive power. Let us assume the lens of 3D glasses has refractive power of $D_C$. In order to keep the focus on $FP_1$, total refractive power of the left eye and 3D glasses lens must remain as $D_O$. From Gullstrand equation for summation of refractive powers of two lenses, we can build a relation between $D_O$, $D_V$, and $D_C$.

$$D_O = D_V + D_C - D_V D_C d_I \tag{1}$$

In Eq. (1), $d_I$ is the inter-vertex distance between the eye and 3D glasses lens. When Eq. (1) is satisfied, the viewer can obviously watch the picture at $FP_1$ because the total refractive power is $D_O$. On the other hand, he/she does not feel visual discomfort because accommodation of the left eye becomes $D_V$ keeping the A-V coupling unbroken. The refractive power $D_C$ of 3D glasses compensates the mismatch between $D_O$ and $D_V$. This is compensation of the AVM by refractive power of eye-glasses lens.

The required refractive power $D_C$ can easily be calculated from Eq. (1) as a function of the distances $d_O$, $d_V$, and $d_I$.

$$D_C = \frac{D_O - D_V}{1 - D_V d_I} = \frac{\frac{1}{d_O} - \frac{1}{d_V}}{1 - \frac{1}{d_V} d_I} = \frac{d_V - d_O}{d_O(d_V - d_I)} \tag{2}$$

Although we considered a case that the virtual 3D image locates at V point in front of the 3D display, Eqs. (1) and (2) are also valid when the V point locates behind the display.

**Focus-Adjustable Eye-Glasses**

The compensation method discussed above is not a new one, but sometimes has been used for watching 3D pictures even from 1800s [8]. It seems impractical for motion pictures because location of the virtual 3D image and $d_V$ continuously change. To apply the compensation method to 3D TV and cinema, eye-glasses lens that can changes its refractive power has been required but imppposible.

Fortunately, focus-adjustable eye-glasses that can be utilized for this compensation method are invented just few years ago [9]. They are composed of liquid crystal panels that electro-optically diffract incident lights pixel by pixel, behaving like optical lenses as a result. By controlling external electric field applied to the pixels, its refractive power and focal length are continuously changeable. We would like to propose that, combining the compensation method and focus-adjustable eye-glasses, we could solve the AVM problem and reduce the visual fatigue. According to Ref. 9, the focus-adjustable eye-glasses were originally developed for presbyopia, but their ability of changing refractive power is yet restricted to a range between -1 and +1 D. It is insufficient for utilization in 3D display system, so that further investigation is needed to enhance the range.

**Estimation of Virtual Image Location**

To make the focus-adjustable 3D glasses work in practice, we have to calculate continuously the refractive power $D_C$ from Eq. (2). Among $d_I$, $d_O$, and $d_V$, we think $d_I$ and $d_O$ are easy to know because they are directly measurable.

In optometry, $d_I$ is fixed to 12 mm during the fitting process, though 3D glasses are not yet fitted by optometrist or optician. Therefore, $d_I$ can be approximately regarded as a constant. If exits, the error of $d_I$ is within few millimeters, while that of $d_V$ is tens of centimeters. Effect of the error of $d_I$ on Eq. (2) may be very small. Existence of the refractive power $D_C$ slightly changes the position of nodal point $N_1$, which must be considered in calculation. However, since $d_O$ and $d_V$ are much longer than the change of $N_1$, their errors are not so effective on Eq. (2) that is supposed to hold good approximately.

As to $d_O$, it can be directly measured by signal communication between the 3D glasses and display. In the shutter-glasses type 3D glasses, 3D display has to continuously communicate with the 3D glasses for synchronization. Electronic measurement of $d_O$ may be possible if using this communication.

We also need $d_V$ to calculate $D_C$. It is very hard to determine the depth perceived by human binocular vision, but geometric determination of the location of V point is relatively easy. Assuming that the 3D display, the 3D glasses, and the viewer's both eyes are aligned parallel to one another as shown in Fig. 3, a proportional relationship given below is valid because triangles V-$FP_1$-$FP_2$ and V-$N_1$-$N_2$ are similar to each other.

$$\frac{d_{FP}}{d_{PD}} = \frac{d_O - d_V}{d_V} \qquad (3)$$

In Eq. (3), $d_{FP}$ is distance between two motion pictures divided for inducing the binocular disparity, and $d_{PD}$ is the inter-pupillary distance between the viewer's both eyes. Now we can calculate $d_V$ from Eq. (4) given below.

$$d_V = \frac{d_O}{1 + \dfrac{d_{FP}}{d_{PD}}} \qquad (4)$$

Eq. (4) is not just a substitution of $d_V$ for $d_{PD}$ and $d_{FP}$. Direct measurement of $d_V$ is almost impossible, but $d_{PD}$ and $d_{FP}$ are measurable quantities. The inter-pupillary distance $d_{PD}$ is an individually fixed parameter, which is about 60 mm on average. Like $d_I$, it can be measured and input by optometrist for private 3D glasses, or approximately regarded as an optometric constant. The distance $d_{FP}$ can be drawn from electronic data processing of 3D display. For example, a 3D display monitor that adjusts the 'depth' by input parameter is commercially available [10], therefore $d_{FP}$ must be a programming parameter that can easily be obtained from display.

**Discussion**

It is supposed that rapid change of the refractive power of eye-glasses can make the viewer feel dizzy. To avoid such dizziness, change of the refractive power of 3D glasses must not be so sharp but smoothened. It is well known in optometry that the A-V coupling is not so strong that small mismatch does not cause visual discomfort. The smoothened change of refractive power and slightly broken A-V coupling would not induce 3D visual fatigue.

Such flexibility of the A-V coupling can also be useful for approximations in our operation scheme. In the above calculations, we adopted approximations of $\boldsymbol{d}_\text{I}$ and $\boldsymbol{d}_\text{PD}$, and assumed that the 3D display, the 3D glasses, and the viewer's both eyes are aligned parallel to one another. These approximations and assumption can cause small error in estimation of $\boldsymbol{D}_\text{C}$, but flexibility of the A-V coupling is expected to buffer the error.

One of possible arguments against our proposal is that there exist multiple objects of different depths within a 3D scene. Even the V point of one object locates in front of the 3D display while that of another object locates behind the 3D display. Which object do we have to focus on? It is related to the visual attention problem in vision science, which is very complicate. Considering the problem just in an instrumental point of view, we think a possible solution is regarding normal direction of the 3D glasses as viewer's gaze. The direction can electronically be detected, and the refractive power is adjusted to the depth of object on the gaze. For more systematic solution, well-developed eye movement tracking technic can be adopted to detect viewer's gaze. Don't the multiple objects with various depths in a field of view cause any confusion? Maybe not. The situation is the same with natural vision, in which many objects exist with various distance in a field of view. The human eyes focus on an object with narrow view angle, so that the multiple objects with various depths in a 3D scene are not supposed to cause any confusion.

Various solutions have been proposed for the AVM problem, and most of them can be classified into two categories, wave front reconstructing or volumetric displays [6]. For the wave front reconstruction, the 3D display is the multi-view type or the holography type. Any solution to the AVM problem has not been considered for recently commercialized 2-view type 3D displays. We think the compensation method using focus-adjustable lens is an expectable solution for the 2-view type 3D displays.

**Concluding Remarks**

We proposed that the AVM in 3D motion pictures can be compensated by refractive power of focus-adjustable 3D glasses. Mathematical equations are developed for calculation of required refractive power of 3D glasses. The compensation method is supposed to remove visual discomfort caused by the AVM. To realize this scheme, further theoretical, instrumental, and clinical studies are needed.

**Figures**

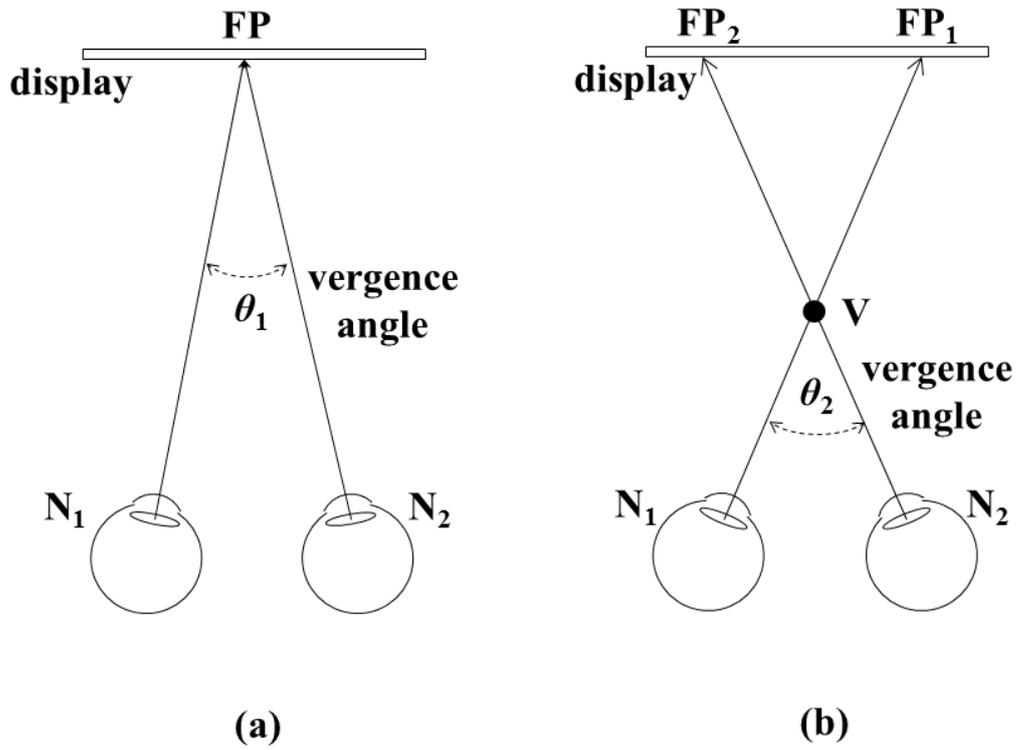

Figure 1. Relations between fixation points, vergence angles, and 3D virtual image when a viewer watches (a) a 2D display or (b) a 2-view type 3D display.

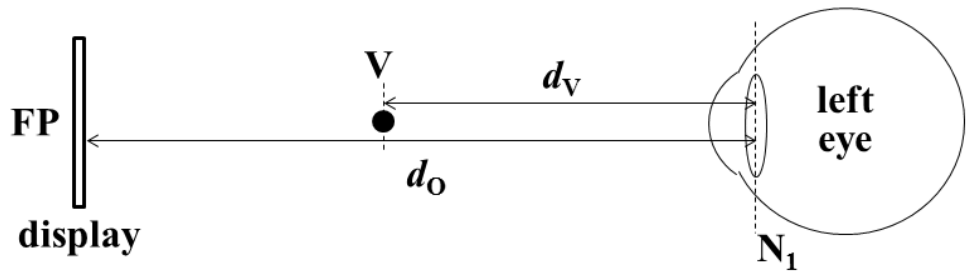

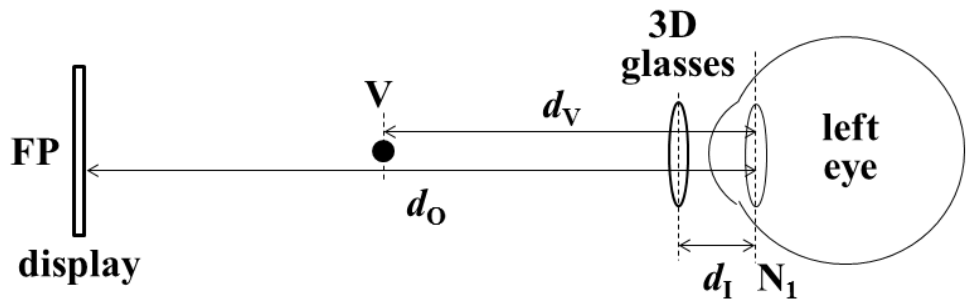

**Figure 2.** Simplified length relations between $d_O$, $d_V$, and $d_I$ when the viewer watches (a) by the naked eye or (b) through a lens of the 3D glasses.

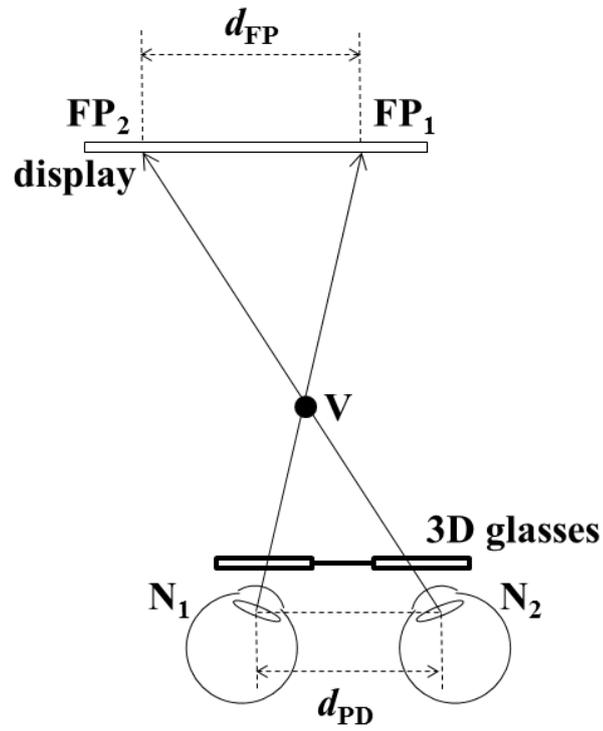

Figure 3. Geometric relations between $d_{PD}$, $d_{FP}$, V-$N_1$(= $d_V$), and V-$FP_1$(= $d_O$−$d_V$). Triangles V-$FP_1$-$FP_2$ and V-$N_1$-$N_2$ are similar to each other.